\documentclass[aps,prl,preprint,groupedaddress,notitlepage]{revtex4}
\usepackage{graphicx}
\usepackage{amsmath}
\graphicspath{{./figs/}}

\preprint{NuFact15 - Rio de Janeiro, Brazil - August 2015}

\begin{document}

%\preprint{NuFact15 - Rio de Janeiro, Brazil - August 2015}

\title{Correlations in neutrino-nucleus scattering}
\thanks{\it Presented at NuFact15, 10-15 Aug 2015, Rio de Janeiro, 
Brazil [C15-08-10.2]}



\author{T.~Van Cuyck}
\email{Tom.VanCuyck@UGent.be}
\thanks{Speaker}
\affiliation{Department of Physics and Astronomy,\\
Ghent University,\\ Proeftuinstraat 86,\\ B-9000 Gent, Belgium}
\author{V.~Pandey}
\email{Vishvas.Pandey@UGent.be}
\affiliation{Department of Physics and Astronomy,\\
Ghent University,\\ Proeftuinstraat 86,\\ B-9000 Gent, Belgium}
\author{N.~Jachowicz}
\email{Natalie.Jachowicz@UGent.be}
\affiliation{Department of Physics and Astronomy,\\
Ghent University,\\ Proeftuinstraat 86,\\ B-9000 Gent, Belgium}
\author{R.~Gonz\'{a}lez-Jim\'{e}nez}
\affiliation{Department of Physics and Astronomy,\\
Ghent University,\\ Proeftuinstraat 86,\\ B-9000 Gent, Belgium}
\author{M.~Martini}
\affiliation{Department of Physics and Astronomy,\\
Ghent University,\\ Proeftuinstraat 86,\\ B-9000 Gent, Belgium}
%\affiliation{ESNT, CEA-Saclay, DSM, IRFU,\\
%Service de Physique Nuclaire, \\ F-91191 Gif-sur-Yvette Cedex, France}
\author{J.~Ryckebusch}
\affiliation{Department of Physics and Astronomy,\\
Ghent University,\\ Proeftuinstraat 86,\\ B-9000 Gent, Belgium}
\author{N.~Van Dessel}
\affiliation{Department of Physics and Astronomy,\\
Ghent University,\\ Proeftuinstraat 86,\\ B-9000 Gent, Belgium}

%  \author{Jane Q. Physicist}
%  \email[]{jphysicist@fnal.gov}
%  %\homepage[]{Your web page}
%  \thanks{Speaker}
%  %\altaffiliation{}
%  \affiliation{Fermilab}
%  \author{John X. Theorist}
%  \email[]{jtheorist@cern.ch}
%  %\homepage[]{Your web page}
%  %\thanks{}
%  %\altaffiliation{}
%  \affiliation{
%  Department of Physics and Astronomy,\\
%  Ghent University,\\
%  Proeftuinstraat 86,\\
%  B-9000 Gent, Belgium
%  }


\date{\today}

\begin{abstract}
  We present a detailed study of  charged-current quasielastic neutrino-nucleus scattering and of the influence of correlations on one- and two-nucleon knockout processes.
The quasielastic neutrino-nucleus scattering cross sections, including the influence of long-range correlations, are evaluated within a continuum random phase approximation approach.  The short-range correlation formalism is implemented in the impulse approximation by shifting the complexity induced by the correlations from the wave functions to the operators.
The model is validated by confronting $(e,e^\prime)$ cross-section predictions with electron scattering data in the kinematic region where the quasielastic channel is expected to dominate. Further, the $^{12}$C$(\nu_\mu,\mu^-)$ cross sections relevant for neutrino-oscillation experiments are studied.  Double differential $^{12}$C$(\nu_\mu,\mu^-)$ cross sections,  accounting for long- and short-range correlations in the one-particle emission channel and short-range correlations in the two-particle two-hole channel, are presented for
  kinematics relevant for recent neutrino-nucleus scattering measurements.
\end{abstract}

%\keywords{}

\maketitle

\section{Introduction}
One of the major issues in accelerator-based neutrino-oscillation experiments is the need for accurate predictions of neutrino-nucleus scattering cross sections at intermediate (0.01 - 2 GeV) energies. A model where the $W$ boson interacts with a single nucleon, which subsequently leaves the residual nucleus unhindered, does not accurately describe recent experimental measurements of neutrino and antineutrino cross sections. A major complication stems from the fact that typical neutrino-nucleus
measurements do not uniquely determine the nuclear final state, as only the energy-momentum of the final muon are measured. In order to explain the discrepancy between theory and experiment, one needs a model that includes nuclear correlations, meson-exchange currents and final-state interactions. 
In this work, we focus on the influence of nuclear correlations on inclusive quasielastic (QE) cross sections. First we will discuss long-range correlations in a continuum random phase approximation (CRPA) approach, and secondly, the influence of short-range correlations (SRCs).\\
The model described below was used successfully in the description of exclusive electron-scattering processes \cite{Ryckebusch:1988aa,Ryckebusch:1989nn}, low-energy and supernova neutrino processes \cite{Jachowicz:1998fn,Jachowicz:2002rr,Jachowicz:2004we}, and extended to the description of inclusive quasielastic electroweak scattering cross sections at intermediate energies in \cite{Pandey:2013cca,Pandey:2014tza}.

\section{quasielastic Neutrino-nucleus scattering cross section}
In this section, we briefly describe the approach for the calculation of the nuclear response for inclusive electron- and neutrino-nucleus scattering in the QE region. Considering electron scattering off a nucleus, the double differential $A(e,e^\prime)$ cross section is written as
  \begin{align}
    \dfrac{\mathrm{d}\sigma}{\mathrm{d}E_{e'} \mathrm{d}\Omega_{e'}} = \left( \frac{\alpha \cos(\theta_{e'}/2)}{2E_e\sin^2(\theta_{e'}/2)}\right)^2 \bigl[& v^e_{L} W_{CC} + v^e_T W_T \bigr],
  \end{align}
  with $\alpha$ the fine-structure constant and $\theta_{e'}$ the scattering angle of the electron. 
  For CC neutrino-nucleus $A(\nu_\mu,\mu^-)$ interactions, the cross section is expressed as
  \begin{align}
    \dfrac{\mathrm{d}\sigma}{\mathrm{d}E_\mu \mathrm{d}\Omega_\mu} = \left( \frac{G_F \cos(\theta_c) E_\mu}{2\pi}\right)^2 \zeta \bigl[& v_{CC} W_{CC} + v_{CL} W_{CL} + v_{LL} W_{LL} + v_T W_T  - v_{T'} W_{T'} \bigr],
  \end{align}
  with $G_F$ the Fermi constant, $\theta_c$ the Cabibbo angle and the kinematic factor $\zeta$
  \begin{align}
    \zeta=\sqrt{1-\frac{m_\mu^2}{E_\mu^2}}.
  \end{align}
  The functions $v$ contain the leptonic information and the $W$ are nuclear response functions, they are defined as products of transition matrix elements $\mathcal{J}_\lambda$
  \begin{align}
    \mathcal{J}_\lambda = \langle \Psi_\textnormal{f} | \widehat{J}_\lambda^\textnormal{nucl} | \Psi_\textnormal{i} \rangle, \label{eq:nuc}
  \end{align}
  with $| \Psi_\textnormal{f} \rangle$ and $| \Psi_\textnormal{i} \rangle$ the final and initial nuclear state and $\widehat{J}_\lambda^\textnormal{nucl}$ the spherical components of the nuclear current. The expressions for the $v$ and $W$ can be found in Ref.~\cite{Pandey:2014tza}.

\section{Hartree-Fock mean field model} 
A key element in the model presented here is the non-relativistic impulse approximation. The Hartree-Fock (HF) single-particle bound-states and the continuum wave functions are obtained by solving the Schr\"odinger equation using an effective Skyrme interaction.   The SkE2 Skyrme parameterization is based on a fit to ground-state and low-lying excited state properties of spherical nuclei \cite{Ryckebusch:1988aa,Ryckebusch:1989nn}.
The fact that the outgoing nucleon's wave function is generated in a (real) nuclear potential partially includes final-state interactions, in a natural way.  The influence of the  spreading width of the particle states is taken into account by a folding procedure \cite{Pandey:2014tza}.
The impact of the Coulomb potential of the nucleus on the outgoing lepton is implemented using a modified effective momentum approach (MEMA) \cite{Engel:1997fy}.  As the description of the nuclear dynamics is non-relativistic, relativistic corrections are implemented based on the effective scheme proposed in \cite{Jeschonnek:1997dm}.

\section{Long-range correlations}
\begin{figure}
   \includegraphics[width=\textwidth,trim=0cm 1cm 0cm 0.5cm, clip=true]{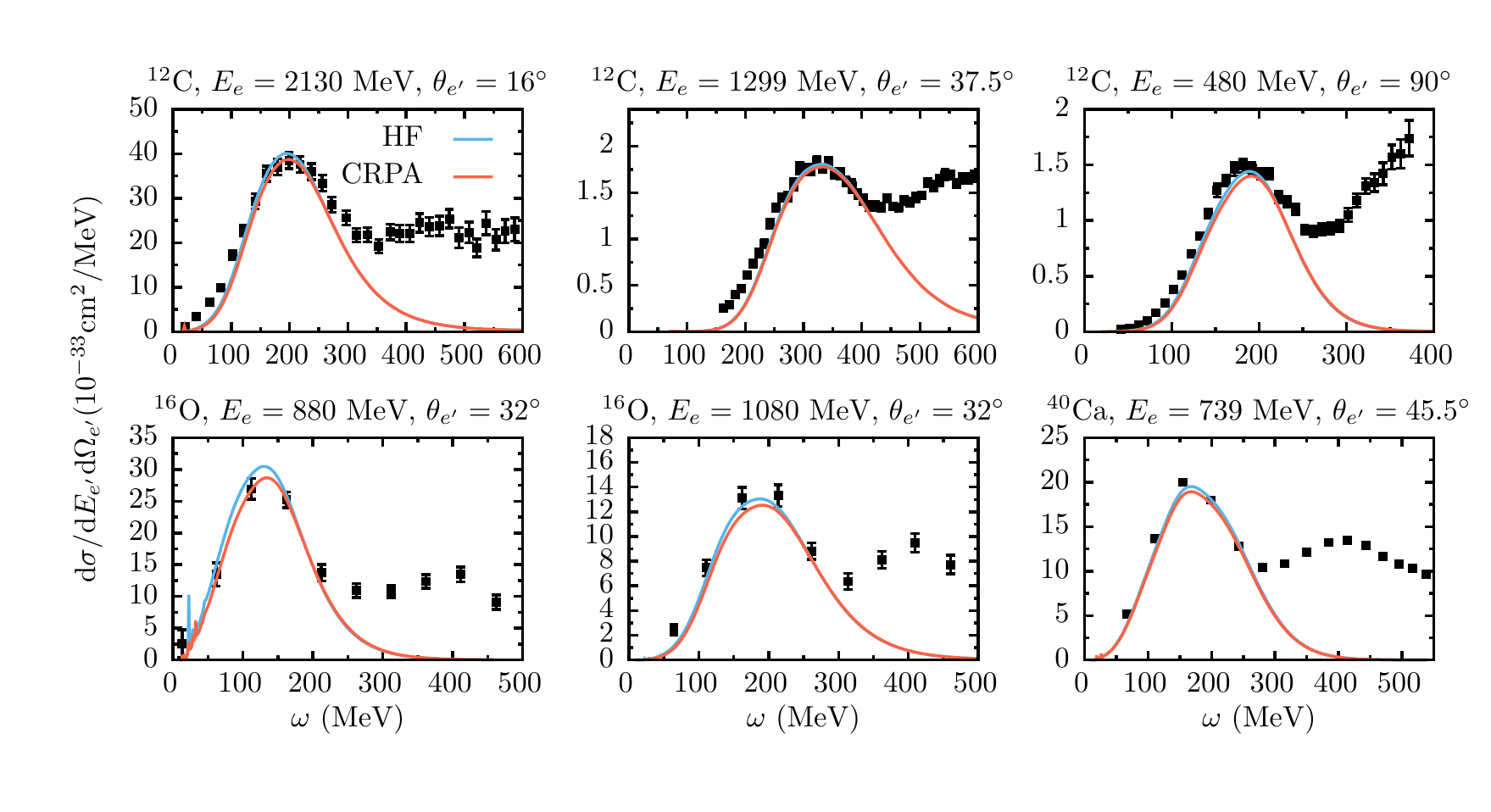}
   \caption{Double differential $(e,e^\prime)$ cross section with $^{12}$C, $^{16}$O and $^{40}$Ca as target nuclei. Data are from Refs.~\cite{Bagdasaryan:1988aa,Sealock:1989nx,Barreau:1983ht,Anghinolfi:1996vm,Williamson:1997zz}.}\label{fig:1}
 \end{figure}

Long-range correlations are introduced using a continuum random phase approximation approach.  The CRPA is based on a Green's function formalism, where the CRPA propagator is obtained by the iteration to all orders of the first-order contribution to the particle-hole Green's function
\begin{align}
  \Pi^{(RPA)}(x_1,x_2;E_x) &= \Pi^{(0)}(x_1,x_2;E_x)\nonumber \\ & + \frac{1}{\hbar} \int \mathrm{d} x \mathrm{d} x^\prime \Pi^{(0)} (x_1,x;E_x) \tilde{V}(x,x^\prime) \Pi^{(RPA)}(x^\prime,x_2;E_x),
\end{align}
with $\tilde{V}$ the antisymmetrized Skyrme residual interaction.
The same Skyrme SkE2 parameterization that is used to generate the HF single-particle wave functions is  used as $ph$-interaction in the RPA calculation, assuring consistency of the formalism with regards to the nucleon interaction that is used.
The $Q^2$ running of the residual interaction is controled by a dipole form factor at the nucleon vertices \cite{Pandey:2014tza}.  
The CRPA wave-functions $| \Psi^{RPA}_C \rangle$ and transition densities are then related to the unperturbed wave-functions $| ph^{-1} \rangle$ through
%\begin{align}
%  \left|\Psi_C^{RPA}(E)\right\rangle &= \left|ph^{-1}(E)\right\rangle + \int\! \textnormal{d}x_1\! \int\! \textnormal{d}x_2\;\; \widetilde{V}(x_1,x_2) \nonumber \\ & \hspace{-1.5cm}
%  \sum_{C'} {\cal P}  \int \textnormal{d} \varepsilon_{p'} \Biggl[\; \frac{\psi_{h'}(x_1)
%    \psi^{\dagger}_{p'}(x_1,\varepsilon_{p'}) }{E-\varepsilon_{p'h'}}\left|p'h'^{-1}(\varepsilon_{p'h'})\right\rangle \nonumber \\ & \hspace{0.85cm}- \frac{\psi_{h'}^{\dagger}(x_1)
%\psi_{p'}(x_1,\varepsilon_{p'})}{E+\varepsilon_{p'h'}}\left|h'p'^{-1}(-\varepsilon_{p'h'})\right\rangle \Biggr] 
%\left\langle\Psi_0\left|\hat{\psi}^{\dagger}(x_2)\hat{\psi}(x_2)\right|\Psi_C(E)\right\rangle .\label{psirpa} 
%\end{align}
%Or, written as a superposition of particle-hole and hole-particle states,
\begin{align}
  | \Psi^{RPA}_C \rangle &= \sum_{C'} \left[ X_{C,C^\prime} | p^\prime h^{\prime -1} \rangle - Y_{C,C^\prime} | h^\prime p^{\prime -1} \rangle \right],
\end{align}
with
\begin{eqnarray} 
 X_{C,C'}(E,\varepsilon_{p'})\!&=&\!\delta_{C,C'}\;\delta(E-\varepsilon_{p'h'})  
 +\;\;{\cal P} \int \!\textnormal{d}x_1\!\int \!\textnormal{d}x_2 \;\:\widetilde{V}(x_1,x_2)\;\nonumber \\ &&\hspace*{1.5cm}\frac{\psi_{h'}(x_1)
\psi^{\dagger}_{p'}(x_1,\varepsilon_{p'}) }{E-\varepsilon_{p'h'}}\:
\left\langle\Psi_0\left|\hat{\psi}^{\dagger}(x_2)\hat{\psi}(x_2)\right|\Psi_C(E)\right\rangle\, ,
\end{eqnarray}
and
\begin{eqnarray}
  Y_{C,C'}(E,\varepsilon_{p'})\!&=&\!\int \!\textnormal{d}x_1\!\int \!\textnormal{d}x_2 \;\:\widetilde{V}(x_1,x_2)\;\nonumber \\ &&\hspace*{1.5cm}\frac{\psi^{\dagger}_{h'}(x_1)
\psi_{p'}(x_1,\varepsilon_{p'}) }{E+\varepsilon_{p'h'}}\:
\left\langle\Psi_0\left|\hat{\psi}^{\dagger}(x_2)\hat{\psi}(x_2)\right|\Psi_C(E)\right\rangle\, ,\label{Y}
 \end{eqnarray}
\noindent
with $C$ denoting all quantum numbers representing an accessible channel. These equations reflect the fact that RPA wave functions are a superposition of $ph$- and $hp$-excitations out of a correlated ground state.\\
In FIG.~\ref{fig:1}, the HF and CRPA predictions are compared with double-differential electron-scattering data for three different target nuclei. In view of the fact that our description only considers the QE channel, while the measurements include contributions such as $\Delta$ excitations and two-particle knockout, our numerical calculations provide a fair agreement with the data in the kinematic range presented here. The difference between the HF and CRPA results are
sizable for $Q^2 \le 0.25$ (GeV/c)$^2$, see Ref.~\cite{Pandey:2014tza}. For the results presented here, which account for higher $Q^2$ values, the HF and CRPA cross sections are comparable.

\section{Short-Range Correlations}
\begin{figure}
  \includegraphics[width=0.55\textwidth]{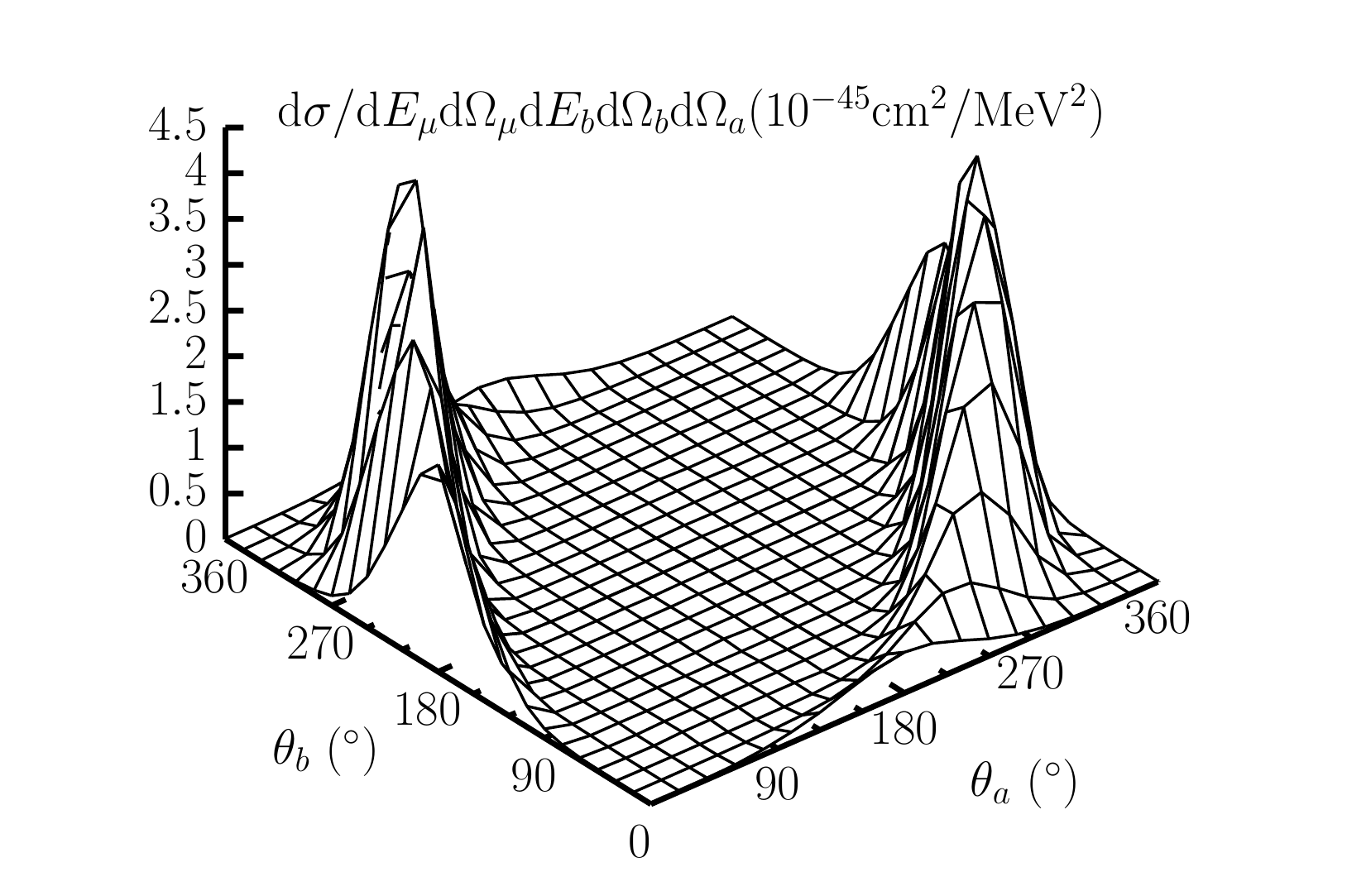}%
  \caption{Exclusive $^{12}$C$(\nu_\mu,\mu^- N_a N_b)$ cross section at $E_{\nu_\mu} = 750$ MeV, $E_\mu = 550$ MeV, $\theta_\mu = 15^\circ$ and $T_\textnormal{p} = 50$ MeV with outgoing nucleons in the lepton scattering plane.}\label{fig:excl}
\end{figure}

To account for SRCs in neutrino-nucleus scattering, we rely on a model developed for exclusive as well as semi-exclusive electron-nucleus scattering cross sections \cite{VanderSluys:1995rp,Ryckebusch:1997gn,Janssen:1999xy,Vanhalst:2014cqa}. This work is a first step in the extension of this model towards the weak sector.\\
The correlated wave functions $| \Psi \rangle$ are constructed by applying a many-body correlation operator $\widehat{\mathcal{G}}$ to the uncorrelated wave functions $| \Phi \rangle$
\begin{align}
  | \Psi \rangle = \frac{1}{\sqrt{\mathcal{N}}} \widehat{\mathcal{G}} | \Phi \rangle,
\end{align}
with $\mathcal{N} = \langle \Phi | \widehat{\mathcal{G}}^\dagger \widehat{\mathcal{G}} | \Phi \rangle$ the normalization constant. In the construction of the correlation operator, one is guided by the features of the one-boson exchange nucleon-nucleon force. In this work, only the central and tensor part of the correlation operator are considered, spin-isospin correlations will be included in future work,
\begin{align}
  \widehat{\mathcal{G}} = \widehat{\mathcal{S}} \left( \prod_{i<j}^A \left[ 1 + \widehat{l}(i,j) \right] \right),
\end{align}
with
\begin{align}
  \widehat{l}(i,j) &= -\widehat{g}(i,j) + \widehat{t}(i,j) \\
  &=-g_c(r_{ij}) + f_{t\tau}(r_{ij}) \widehat{S}_{ij} \left(\vec{\tau}_i \cdot \vec{\tau}_j \right),
\end{align}
where $\widehat{S}$ is the symmetrization operator, $\widehat{S}_{ij}$ the tensor operator $\frac{3}{r_{ij}^2} \left( \vec{\sigma}_i \cdot \vec{r}_{ij} \right) \left(\vec{\sigma}_j \cdot \vec{r}_{ij} \right) - \left( \vec{\sigma}_i \cdot \vec{\sigma}_j \right)$, $g_c(r_{ij})$ the central correlation function and $f_{t\tau}(r_{ij})$ the tensor correlation function. In the calculations presented in this work we used the central correlation function parameterization by Gearhaert and Dickhoff
\cite{gearhaert} and the tensor correlation function by Pieper \textit{et al.}~\cite{Pieper:1992gr}. Ref.~\cite{Vanhalst:2012ur} provides arguments and evidence to support the fact that these correlation functions can be considered realistic.\\
When calculating transition matrix elements between correlated states $| \Psi \rangle$, one can shift the effect of the correlations towards the transition operators and calculate matrix elements between uncorrelated states $| \Phi \rangle$ with an effective transition operator. In the IA, the many-body nuclear current operator $\widehat{J}_\lambda^\textrm{nucl}$ can be written as a sum of one-body currents $\widehat{J}_\lambda^{[1]}(i)$. To account for SRCs, the current in
Eq.~\ref{eq:nuc} is replaced with an effective current
\begin{align}
  \langle \Psi_\textnormal{f} | \widehat{J}_\lambda^\textrm{nucl} | \Psi_\textnormal{i} \rangle = \frac{1}{\mathcal{N}} \langle \Phi_\textnormal{f} | \widehat{\mathcal{G}}^\dagger \widehat{J}_\lambda^\textrm{nucl} \widehat{\mathcal{G}} | \Phi_\textnormal{i} \rangle = \frac{1}{\mathcal{N}} \langle \Phi_\textnormal{f} | \widehat{J}_\lambda^{\rm{eff}} | \Phi_\textnormal{i} \rangle,
\end{align}
with
\begin{align}
  \widehat{J}^{\rm{eff}}_\lambda &= \left( \prod^A_{j<k} \left[ 1 +\widehat{l}(j,k)  \right] \right)^\dagger \sum_{i=1}^A \widehat{J}_\lambda^{[1]}(i)  \left( \prod^A_{l<m} \left[ 1 +\widehat{l}(l,m)  \right] \right).
\end{align}
Relying on the short-range behavior of the correlations, the effective current is approximated as 
\begin{align}
  \widehat{J}_\lambda^{\rm{eff}} &\approx \sum_{i=1}^A \widehat{J}_\lambda^{[1]}(i) +\sum_{i<j}^A \widehat{J}_\lambda^{[1],\rm{in}}(i,j) + \left[ \sum_{i<j}^A \widehat{J}_\lambda^{[1],\rm{in}}(i,j) \right]^\dagger,
\end{align}
where the first term is the nuclear current in the IA, and the second term is a two-body current which is the product of a one-body current and a correlation operator
\begin{align}
  \widehat{J}_\lambda^{[1],\rm{in}}(i,j) &= \left[ \widehat{J}_\lambda^{[1]}(i) + \widehat{J}_\lambda^{[1]}(j) \right] \widehat{l}(i,j).
\end{align}
This model is used to study the effect of SRCs on the quasielastic double differential neutrino-nucleus scattering cross section. Due to the two-body structure of the effective operator, the SRCs influence the $1p1h$ as well as the $2p2h$ channel. \\
FIG.~\ref{fig:excl} shows the result of an exclusive cross section calculation. A striking feature of the displayed cross section is the dominance of back-to-back nucleon knockout, reminiscent of the 'hammer events' seen by the ArgoNeuT collaboration \cite{Acciarri:2014gev}. This feature is independent of the interacting lepton or the type of
two-body interaction \cite{Ryckebusch:1993tf,Ryckebusch:1997gn}. \\
The contribution of the $2p2h$ channel to the double differential cross section involves an integration over the phase space of the undetected nucleons as
outlined in Refs.~\cite{Ryckebusch:1993tf,VanderSluys:1995rp} 
\begin{align}
  \frac{\textrm{d} \sigma}{\textrm{d} E_\mu \textrm{d} \Omega_\mu } (\nu_\mu,\mu^-) = \int \textrm{d} T_b \textrm{d} \Omega_b \textrm{d} \Omega_a
  \frac{\textrm{d} \sigma}{\textrm{d} E_\mu \textrm{d} \Omega_\mu \textrm{d} T_b \textrm{d} \Omega_b \textrm{d} \Omega_a} (\nu_\mu,\mu^- N_a N_b) .
\end{align}

\begin{figure}
   \includegraphics[width=\textwidth,trim=0cm 0.5cm 0cm 0.2cm,clip=true]{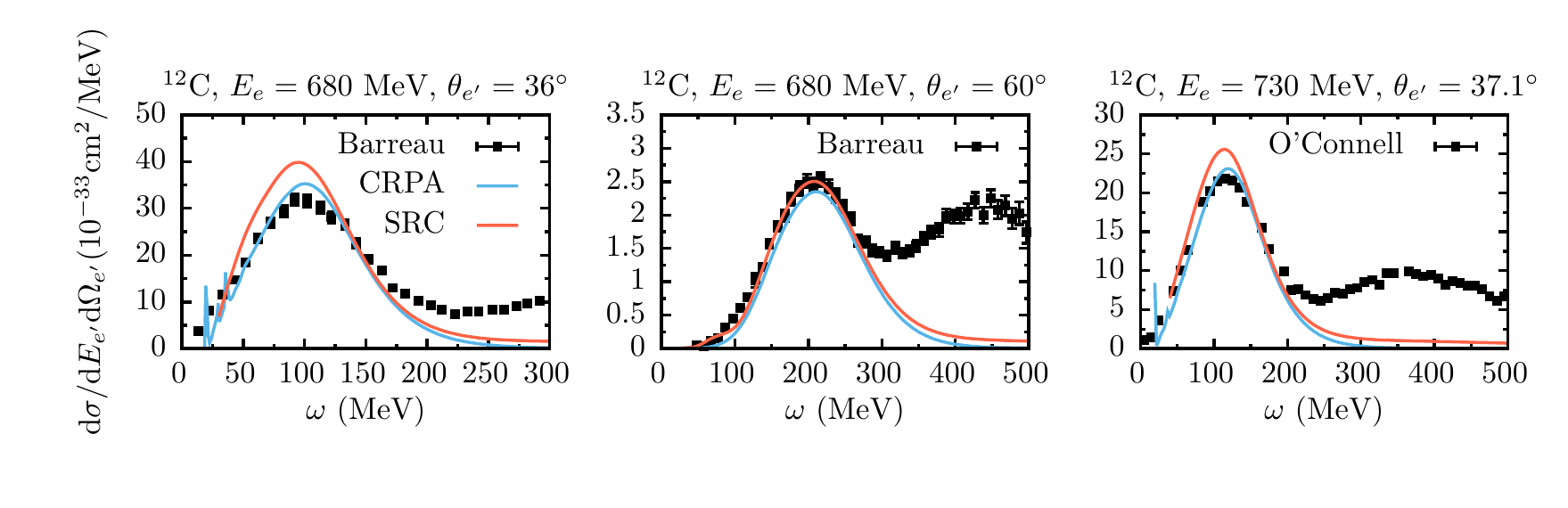}
   \vspace{-1.5cm}
   \caption{Double differential $^{12}$C$(e,e^\prime)$ cross section for three kinematics. Data are from Refs.~\cite{Barreau:1983ht,OConnell:1984nw}.}\label{fig:3}
 \end{figure}

\begin{figure}
   \includegraphics[width=\textwidth,trim=0cm 0.5cm 0cm 0.2cm,clip=true]{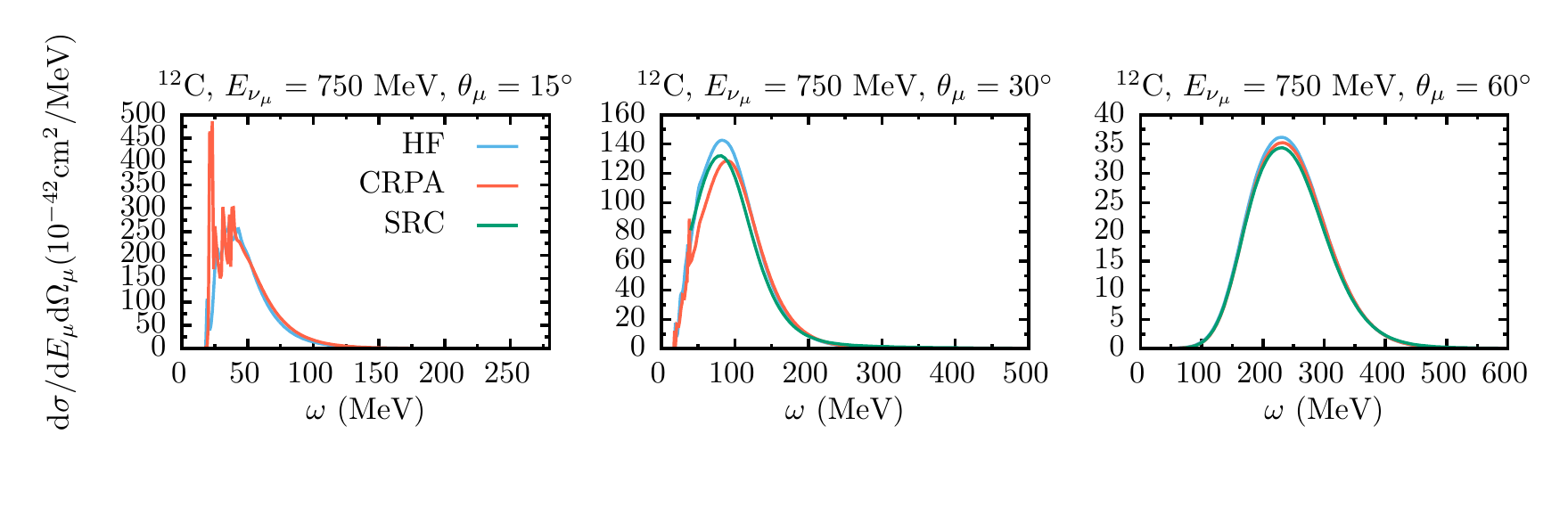}
   \vspace{-1.5cm}
   \caption{Double differential $^{12}$C$(\nu_\mu,\mu^-)$ cross section for three kinematics.}\label{fig:4}
 \end{figure}

\noindent In FIG.~\ref{fig:3}, double differential CRPA $^{12}$C$(e,e^\prime)$ calculations are compared with the model including SRCs in the $1p1h$ and $2p2h$ channels. The CRPA suppression in the QE-region is visible as well as the increase of the cross section in the dip-region due to the two-particle knockout of short-range correlated pairs. FIG.~\ref{fig:4} compares the influence of long- and short-range correlations, accounting for one- and two-particle knockout, on the mean-field $^{12}$C$(\nu_\mu,\mu^-)$ cross section, for three kinematics relevant for accelerator-based neutrino-oscillation experiments. Both models result in a decrease
of the cross section at the QE-peak. 

\section{Summary}
We have presented a discussion of long- and short-range correlations in quasielastic charged-current neutrino-nucleus scattering. We confronted our numerical results with double-differential inclusive $(e,e^\prime)$ electron scattering data and calculated double differential $(\nu_\mu,\mu^-)$ neutrino-nucleus scattering cross sections at energies relevant for recent measurements. A fair agreement with electron-scattering data was reached in the region where the quasielastic channel
is expected to dominate. Furthermore, the framework allows for the prediction of exclusive cross sections, which might provide deeper insight in neutrino experiments detecting the nuclear final state.

\appendix
%\section{}

\begin{acknowledgments}
   This work was supported by the Interuniversity Attraction Poles Programme P7/12 initiated by the Belgian Science Policy Office and the Research Foundation Flanders (FWO-Flanders). The computational resources (Stevin Supercomputer Infrastructure) and services used in this work were provided by Ghent University, the Hercules Foundation and the Flemish Government.
\end{acknowledgments}

% Create the reference section using BibTeX
\bibliographystyle{biblio-persrev}
\bibliography{biblio}

\end{document}